\documentclass[aip,graphicx]{revtex4-1}

\usepackage{xcolor}
\usepackage{xfrac}

\begin{document}

\title{Interface-Selective Extreme Ultraviolet Second Harmonic Generation Away from Linear Absorption Edges}

\author{Bailey R. Nebgen}
\affiliation{Department of Chemistry, University of California Berkeley, Berkeley CA, 94720, USA}
\affiliation{Materials Sciences Division, Lawrence Berkeley National Laboratory, Berkeley CA, 94720, USA}

\author{Shane W. Devlin}
\affiliation{Chemical Sciences Division, Lawrence Berkeley National Laboratory, Berkeley CA, 94720, USA}
\affiliation{Department of Physics, University of Nevada Las Vegas, Las Vegas NV, 89154, USA}

\author{Franky Bernal}
\affiliation{Department of Chemistry, University of California Berkeley, Berkeley CA, 94720, USA}
\affiliation{Chemical Sciences Division, Lawrence Berkeley National Laboratory, Berkeley CA, 94720, USA}

\author{Aga Ridhova}
\affiliation{Department of Electrical and Electronic Engineering, University of Melbourne, Melbourne, Victoria 3010, Australia}

\author{James Bullock}
\affiliation{Department of Electrical and Electronic Engineering, University of Melbourne, Melbourne, Victoria 3010, Australia}

\author{Daniel R. Schacher}
\affiliation{Department of Physics, University of Nevada Las Vegas, Las Vegas NV, 89154, USA}
\affiliation{Department of Physics, Stockholm University, AlbaNova University Center, 106 91 Stockholm, Sweden}

\author{Jake D. Koralek}
\affiliation{Linac Coherent Light Source, SLAC National Accelerator Laboratory, Menlo Park, CA 94025, USA}

\author{Yuya Kubota}
\affiliation{Institute for Solid State Physics, The University of Tokyo, Kashiwa 277-8581, Japan}
\affiliation{RIKEN SPring-8 Center, 1-1-1 Kouto, Sayo, Hyogo 679-5148, Japan}

\author{Gota Yamaguchi}
\affiliation{Institute for Solid State Physics, The University of Tokyo, Kashiwa 277-8581, Japan}
\affiliation{Paul Scherrer Institute, 5232 Villigen PSI, Switzerland}

\author{Shigeki Owada}
\affiliation{RIKEN SPring-8 Center, 1-1-1 Kouto, Sayo, Hyogo 679-5148, Japan}
\affiliation{Japan Synchrotron Radiation Research Institute (JASRI), 1-1-1 Kouto, Sayo, Hyogo 679-5198, Japan}

\author{Toshihide Sumi}
\affiliation{Institute for Solid State Physics, The University of Tokyo, Kashiwa 277-8581, Japan}

\author{XingYu Su}
\affiliation{Institute for Solid State Physics, The University of Tokyo, Kashiwa 277-8581, Japan}

\author{Iwao Matsuda}
\affiliation{Institute for Solid State Physics, The University of Tokyo, Kashiwa 277-8581, Japan}

\author{Keith V. Lawler}
\email{keith.lawler@unlv.edu}
\affiliation{Department of Physics, University of Nevada Las Vegas, Las Vegas NV, 89154, USA}

\author{Walter S. Drisdell}
\email{wsdrisdell@lbl.gov}
\affiliation{Chemical Sciences Division, Lawrence Berkeley National Laboratory, Berkeley CA, 94720, USA}

\author{Michael W. Zuerch}
\email{mwz@berkeley.edu}
\affiliation{Department of Chemistry, University of California Berkeley, Berkeley CA, 94720, USA}
\affiliation{Materials Sciences Division, Lawrence Berkeley National Laboratory, Berkeley CA, 94720, USA}

\author{Craig P. Schwartz}
\email{craig.schwartz@unlv.edu}
\affiliation{Department of Physics, University of Nevada Las Vegas, Las Vegas NV, 89154, USA}
\affiliation{X-ray Sciences Division, Argonne National Laboratory, Argonne IL, 60439 USA}

\begin{abstract}

Extreme ultraviolet (XUV) second harmonic generation (SHG) combines interface selectivity with core-level element specificity but has so far been limited to linear absorption edges. Here, we measure XUVSHG from an indium tin oxide capped MoO$_3$/SiO$_2$/Si photovoltaic heterojunction for fundamental energies of 60--76\,eV, below the Si L$_{2,3}$-edge. Where linear absorption is a featureless valence background, we observe a strong, structured XUVSHG response with maxima near 70 and 76\,eV. Layer-resolved first-principles calculations reproduce both maxima and assign them to the Si/SiO$_2$ interface, the 76\,eV peak to the bridging Si monolayer and the 70\,eV peak to the near-interface substrate. Bethe--Salpeter calculations for bulk Si connect the signal to Si 2$p$ excitons at 140--160\,eV. At the Si L$_{2,3}$-edge, the calculated response is instead governed by MoO$_3$-induced asymmetry at the oxide interface. These findings establish off-edge XUV SHG as a powerful approach for selectively probing distinct buried interfaces within complex material heterostructures.

\end{abstract}

\maketitle

\section{Introduction}

Silicon-based technology is ubiquitous in modern society, underlying much of the semiconductor and photovoltaic industries.\cite{omara1990handbook,vivien2013handbook}
Typically, solar cells contain a p-n junction formed by two layers of Si with different doping that facilitates the separation of electrons and holes at the interface to generate current.\cite{horio2023detecting}
However, recombination losses at the contacts and lack of charge separation remain a limit on solar cell efficiency.
New solar cell architectures are being developed to more efficiently separate electrons and holes at the p-n junction interface.
These include heterojunctions containing new thin-film material layers such as passivated contacts or carrier-selective contacts, which collectively passivate the crystalline Si surface and provide a strong asymmetry in conductivity for electrons and holes.
The function of these material stacks is therefore dictated by the electronic structure at the buried solid-solid junction, which is difficult to access directly with most surface-sensitive techniques.

Studying interfaces, particularly buried interfaces, is a challenge for a variety of reasons.\cite{somorjai2010}
The fundamental difficulty is one of scaling.
The number of atoms at a two-dimensional surface grows with the square of the characteristic sample dimension ($D^2$), whereas the number of bulk atoms grows with its cube ($D^3$).
Even for small samples, the bulk contribution therefore far exceeds the surface contribution.\cite{Bare1988,adamson1977physical}
Perhaps the easiest way to solve this is by using a probe with a fundamentally short penetration or escape depth, such that the signal does not escape from far inside the sample.
Indeed, this is used for a variety of probes ranging from scanning probe microscopy, to various reflective probes, to photoemission (including X-ray photoelectron spectroscopy), to various X-ray scattering techniques.\cite{vanderheide2011xray,KRISHNA2022100332,tompkins2005handbook,collins1990ellipsometry,wang2017ir,mirabella1985internal,alsnielsen2011elements,daillant2009xray,ohnesorge1993true,giessibl1995atomic} 
Unfortunately, many of these techniques will not work on buried interfaces, as the top layer will obscure the relatively small signal from the buried interface.
Other techniques have relatively stringent requirements on surface quality, even when using advanced methods to interpret the data.\cite{greco2021neural} 

The most common methods of studying buried interfaces use nonlinear optics, particularly second harmonic generation (SHG) and sum frequency generation (SFG).\cite{shen2023second,boyd2020nonlinear}
These techniques work by combining two photons to create a photon with energy equal to the combined energy of the input photons.  SHG uses two degenerate photons and SFG uses two nondegenerate photons.
Within the electric-dipole approximation, the process of generating a photon is only possible when centrosymmetry is broken,\cite{shen2023second,boyd2020nonlinear, schacher2026surface}
granting SHG and SFG strong interface sensitivity, since centrosymmetry is necessarily broken at surfaces and interfaces.\cite{bloembergen1968optical}
At high enough photon energies the process becomes bulk sensitive, since at short enough wavelength broken centrosymmetry is encountered on the length scale of interatomic distances within the sample.\cite{glover2012xray,shwartz2014xray} Recent measurements and calculations suggest, however, that in the XUV and soft X-ray regime it remains largely surface sensitive.\cite{schacher2026full,schacher2026surface,lam2018soft,Helk2021,Sumi2023} 
With linear probes, the transitions and edge locations are well established in the X-ray and extreme ultraviolet (XUV) regime.\cite{henke1993xray}
At these energies, the transitions occur from the core or semi-core states leading to relatively consistent transition energies for a given element, even when its chemical environment changes.
Previous XUV and soft X-ray SHG and SFG studies have probed edges that have been observed in linear spectroscopy such as the C K-edge, Li K-edge, O K-edge, Si L-edge, and the Ti, Fe, and As M-edges.\cite{Berger2021,lam2018soft,Uzundal,hoffman2025surface,Woodahl2023,horio2023detecting,Helk2021,Sumi2022,Sumi2023}
However, different selection rules exist for nonlinear techniques such as SFG and SHG, opening up new transitions to study in this regime.\cite{Uzundal}

Nonlinear techniques require strong coherent fields, which are achieved with lasers in the visible and infrared.\cite{boyd2020nonlinear}
Due to the lack of strong coherent fields available from synchrotrons and most laboratory-based sources at XUV to X-ray energies, demonstrations of most nonlinear optical phenomena are more difficult in this regime.\cite{alsnielsen2011elements,Helk2021}
Nonlinear phenomena in the XUV to X-ray regime became broadly feasible with the rise of free electron lasers (FELs), which provide the necessary coherence and peak intensity to drive a variety of nonlinear phenomena at XUV to X-ray photon energies.\cite{Beye2013,glover2012xray}

\begin{figure}
\includegraphics[width=\linewidth]{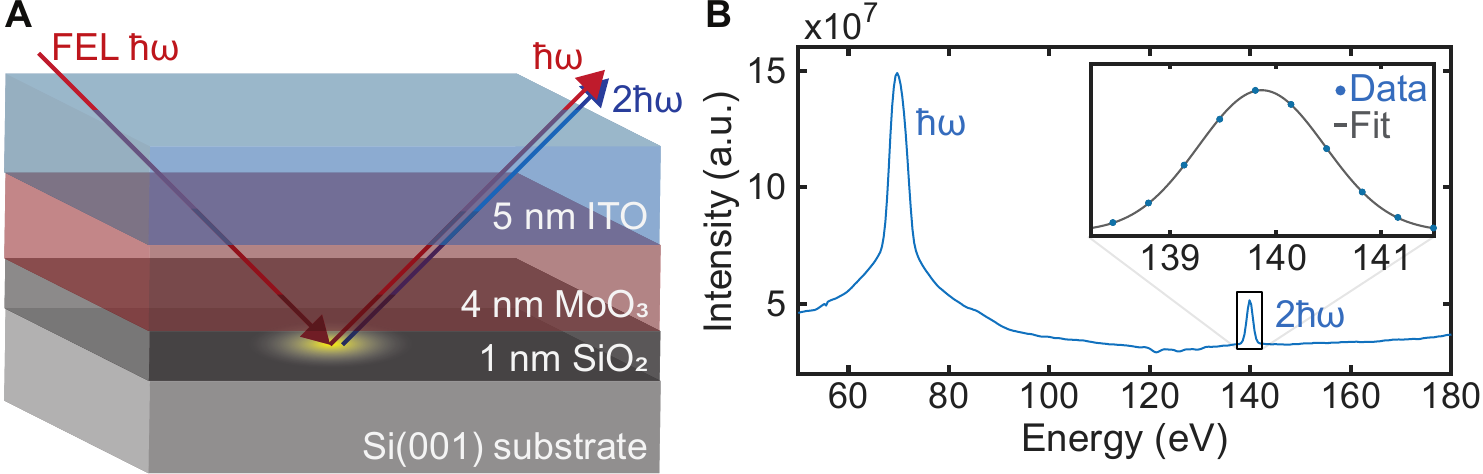}
\caption{Schematic and example of XUV SHG from a MoO$_3$/SiO$_2$/Si solar cell junction. 
\textbf{A} depicts the stacked layers of the solar cell junction with FEL radiation incident at an angle of 45$^\circ$. \textbf{B} shows an example of the FEL fundamental and SH collected simultaneously on the detector after reflection from the sample. The spectrum is a sum of 1000 FEL pulses. The SH is fit to a Gaussian in the inset. \label{fig:schematic}}
\end{figure}

Here we explore the XUV SHG response of buried interfaces in a Si-based solar cell heterojunction containing the hole-conducting layer MoO$_3$ as an example of a promising photovoltaic architecture.\cite{Moxphysicascripta,Bullock2014,BULLOCK2015} We have previously explored the XUV SHG response of Si solar cells with other heterojunctions with excitations at $\sim$100 eV, which is the expected resonance for the Si $L_{2,3}$-edge.\cite{horio2023detecting} 
The full sample stack in this study contains indium tin oxide (ITO), MoO$_3$, SiO$_2$, and a bulk Si wafer substrate (see Fig.~\ref{fig:schematic} \textbf{A}).
This solar cell stack exhibits an unexpected and previously unobserved SHG response in the XUV regime between 60--80\,eV, well separated from direct resonance with the Si $L_{2,3}$-edge at $\sim$100\,eV.
SHG is expected to be strongest from a material when a resonance condition is met either in half-resonance or direct resonance.\cite{shen2023second,schacher2026surface}
In direct resonance, the fundamental photon energy incident on the sample is tuned to be resonant with the target absorption edge and the second harmonic (SH) is detected at twice the energy of the resonance.
In half resonance the fundamental photon energy is tuned to half the energy of the target absorption edge and the SH is detected at the energy of the absorption edge. Our theoretical calculations show that the 60--80\,eV response has a different origin than the $\sim$100\,eV transition. The unexpected XUV SHG response between 60--80\,eV arises from a resonance with a high-energy exciton in Si and has no direct counterpart in a strong linear absorption edge.\cite{henke1993xray}
This new type of XUV SHG response, at least in the case of Si $2p$ excitations between 140 and 160\,eV, makes it possible to probe the nonlinear response of each of several buried interfaces within a complex device independently. It provides access here to the SiO$_2$/Si interface and may also give access to materials whose transitions are not widely 
measurable with other nonlinear techniques.

\section{Results and Discussion}

The experimental stack is shown in Figure~\ref{fig:schematic}\textbf{A}, representing a silicon photovoltaic heterojunction. 
The substrate is a thick Si(001) crystal with a small amount of native oxide ($\sim$1\,nm) on the surface. On top of this substrate is deposited 4\,nm of MoO$_3$ and on top of that 5\,nm of indium tin oxide (ITO).
The FEL is incident on the sample at 45$^\circ$ and passes through all layers of the stack, although it does not transmit through the bulk silicon substrate, which is too thick.
A strong signal at twice the FEL photon energy is observed.
The sum of 1000 shots is shown in Figure~\ref{fig:schematic}\textbf{B}.
The power is varied to generate signal as a function of input intensity, and the resultant second order fit is used to determine $|\chi^{(2)}|^2$.
Using this, an energy-dependent SH in the energy range 60--76\,eV is observed.  

\begin{figure}
\includegraphics[width=\linewidth]{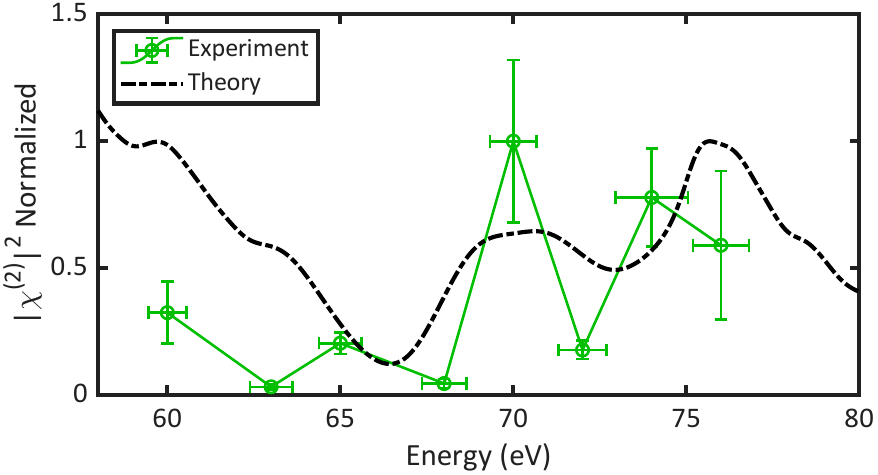}
\caption{XUV SHG spectrum of an MoO$_3$/SiO$_2$/Si solar cell junction. The experimental spectrum is given by the solid green line. Vertical error bars are the standard deviation of $|\chi^{(2)}|^2$ between runs with different filter sets (see Methods). Horizontal error bars represent the standard deviation of the FEL photon energy within each bin. The theoretical modulus squared of the nonlinear susceptibility $|\chi^{(2)}|^2$ for the  MoO$_3$/SiO$_2$/Si simulation stack (depicted in Fig. S1) is given by the dashed line. \label{fig:SHGspectrum}}
\end{figure}

The resulting spectrum is shown in Figure~\ref{fig:SHGspectrum}, with the experiment in green and the theoretical calculation of the $|\chi^{(2)}_{zzz}|^2$ response in the direction normal to the slab face shown as a black dashed line.
While the experimental spectrum is relatively sparse, there is reasonable agreement with the calculated spectrum.
In fact, it is surprising that theory predicts appreciable spectral intensity in that region, as it is not associated with direct or half resonance with a strongly absorbing edge of any material in the photovoltaic stack.
Rather, XUV SHG response in this energy range results from high energy excitons in Si as will be discussed below, differing significantly from XUV SHG at the direct $L_{2,3}$-edge resonance at $\sim$100\,eV.

\begin{figure}
\includegraphics[width=0.9\linewidth]{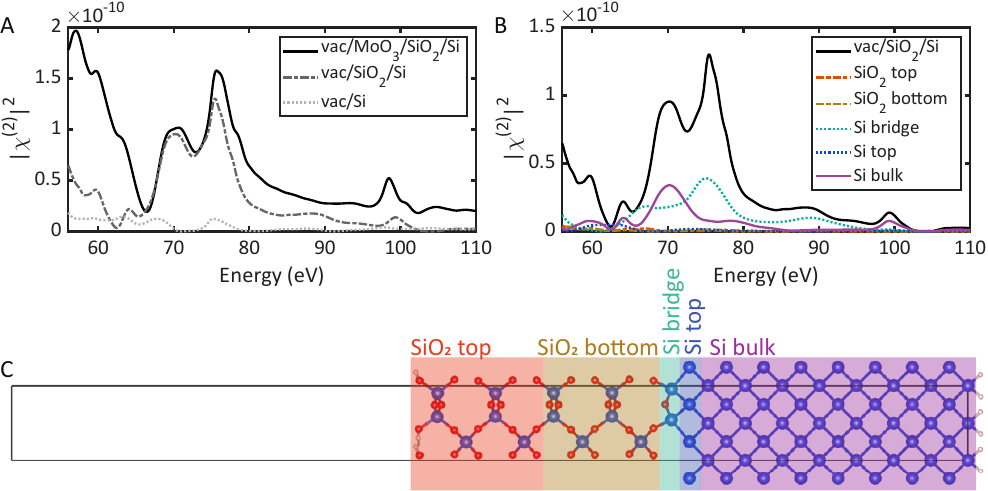}
\caption{Theoretical XUV SHG spectrum of MoO$_3$/SiO$_2$/Si given by modulus squared of nonlinear susceptibility $|\chi^{(2)}|^2$ in the direction normal to the slab face. \textbf{A} shows $|\chi^{(2)}|^2$ for vacuum/MoO$_3$/SiO$_2$/Si (black solid), vacuum/SiO$_2$/Si (grey dash-dot), and vacuum/Si (light grey dotted). \textbf{B} compares the $|\chi^{(2)}|^2$ spectrum for the vacuum/SiO$_2$/Si slab (black) decomposed to the parts of this spectrum, SiO$_2$ top (red dot-dash), SiO$_2$ bottom (brown dot-dash), Si bridge (light blue dot), Si top (dark blue dot), Si bulk (purple solid). The sublayers are labeled on the structure of the full vacuum/SiO$_2$/Si slab in \textbf{C}. \label{fig:SHGspectTheory}}
\end{figure}

While there is strong XUV SHG signal in the spectral range from $\sim$70--76 eV, theoretical calculations of $|\chi^{(2)}|^2$ with different material layers reveal the origin of this unexpected signal.
Figure~\ref{fig:SHGspectTheory}\textbf{A} shows three calculations of $|\chi^{(2)}|^2$ from different material stacks. 
Shown by a black solid line is the MoO$_3$/SiO$_2$/Si simulation stack (depicted in Fig. S1), the same stack without the top MoO$_3$ bilayer (depicted in Fig.~\ref{fig:SHGspectTheory}C) is shown by a gray dash-dot line, and the calculation with only the Si substrate is given by the light gray dotted line.
Critically, around the Si $L_{2,3}$-edge at $\sim$100\,eV, most of the $|\chi^{(2)}|^2$ response is only present in the MoO$_3$/SiO$_2$/Si calculation.
This is the expected result because incident photons at this energy are directly resonant with the Si $L_{2,3}$-edge, and in a previous measurement no significant response was seen for a bare SiO$_2$/Si sample in this energy region.\cite{horio2023detecting}
Therefore, the response at the Si $L_{2,3}$-edge is being driven by the change in electronic structure at the buried interface caused by the MoO$_3$ bilayer.

In contrast, the unexpected XUV SHG signal at $\sim$70\,eV does not decrease significantly when the MoO$_3$ bilayer is removed from the calculation.
This indicates that the XUV SHG response at 70\,eV can be explained in absence of the MoO$_3$ bilayer, and therefore does not primarily originate from the presence of MoO$_3$.
However, when the SiO$_2$ layer is also removed, the $|\chi^{(2)}|^2$ decreases by an order of magnitude. 
Therefore, in our calculations the XUV SHG at $\sim$70\,eV originates from the interface region between SiO$_2$ and Si.
Photons at 70\,eV are able to penetrate deeper into the Si-containing layers of the photovoltaic stack than those at 100\,eV because they are not directly resonant with a strong absorption edge of Si, providing a way to access the $|\chi^{(2)}|^2$ response from additional buried interfaces in the solar cell.

By decomposing the total signal and calculating the response of given layers as illustrated in Figure~\ref{fig:SHGspectTheory}\textbf{C}, the underlying cause of the spectral features in the 60--80\,eV range becomes clearer.
The vacuum/SiO$_2$/Si spectrum from Figure~\ref{fig:SHGspectTheory}\textbf{A} is shown again in Figure~\ref{fig:SHGspectTheory}\textbf{B} (black solid line), and the spectra for the given layers coming from the Si $2p$ states are colored according to the shading in Figure~\ref{fig:SHGspectTheory}\textbf{C}.
The bulk silicon (Si bulk in Fig.~\ref{fig:SHGspectTheory}\textbf{C}), which does not directly touch the oxide but is close enough to the interface to contribute significantly to symmetry breaking, drives the lower energy peak at $\sim$70\,eV.
The peak at $\sim$76\,eV is caused by the silicon atoms that are bridging the bulk and the oxide, which directly interact with the native oxide.
Even though there are fewer of these atoms their cross section is significantly stronger due to the strong symmetry breaking in this interfacial layer.
The other silicon atoms in the oxide do not contribute particularly strongly to the spectrum. Excitations from the Si $2s$ states also contribute little, despite being at half-resonance conditions near the 76\,eV peak. 
The addition of the MoO$_3$ bilayer primarily lowers the contribution of the bulk Si atoms to the 70 and 76\,eV peaks  (plot in supplemental), further emphasizing the role of the strong symmetry breaking in the bridging layer.
Although the Mo $N$-edges have direct and half-resonance conditions in the 60--80\,eV window, only excitations from the Mo $4p$ states produce a non-negligible $|\chi^{(2)}|$ response just below 70\,eV (i.e. half-resonance). This is a minority contribution to the spectrum compared to the bridging or bulk Si atoms.
Notably, unlike in the more molecular case,\cite{hoffman2025surface} the cross terms in the modulus squared have a non-negligible contribution here. 
The decomposition of $|\chi^{(2)}_{zzz}|$, which more clearly sums to the total as opposed to the modulus squared, is shown in the supplemental Figures S2 and S3. 
This analysis of atomic layer contributions demonstrates that the cause of the unexpected XUV SHG at $\sim$70\,eV is due to the silicon and its native oxide and is only perturbed by the coatings.
This provides a convenient spectral range to examine for future studies on the wealth of silicon photovoltaics.

\begin{figure*}[t]
  \centering
  \begin{minipage}[c]{0.65\textwidth}
    \centering
    \makebox[\linewidth][l]{\textbf{A}}\\[0.5mm]
    \includegraphics[width=\linewidth]{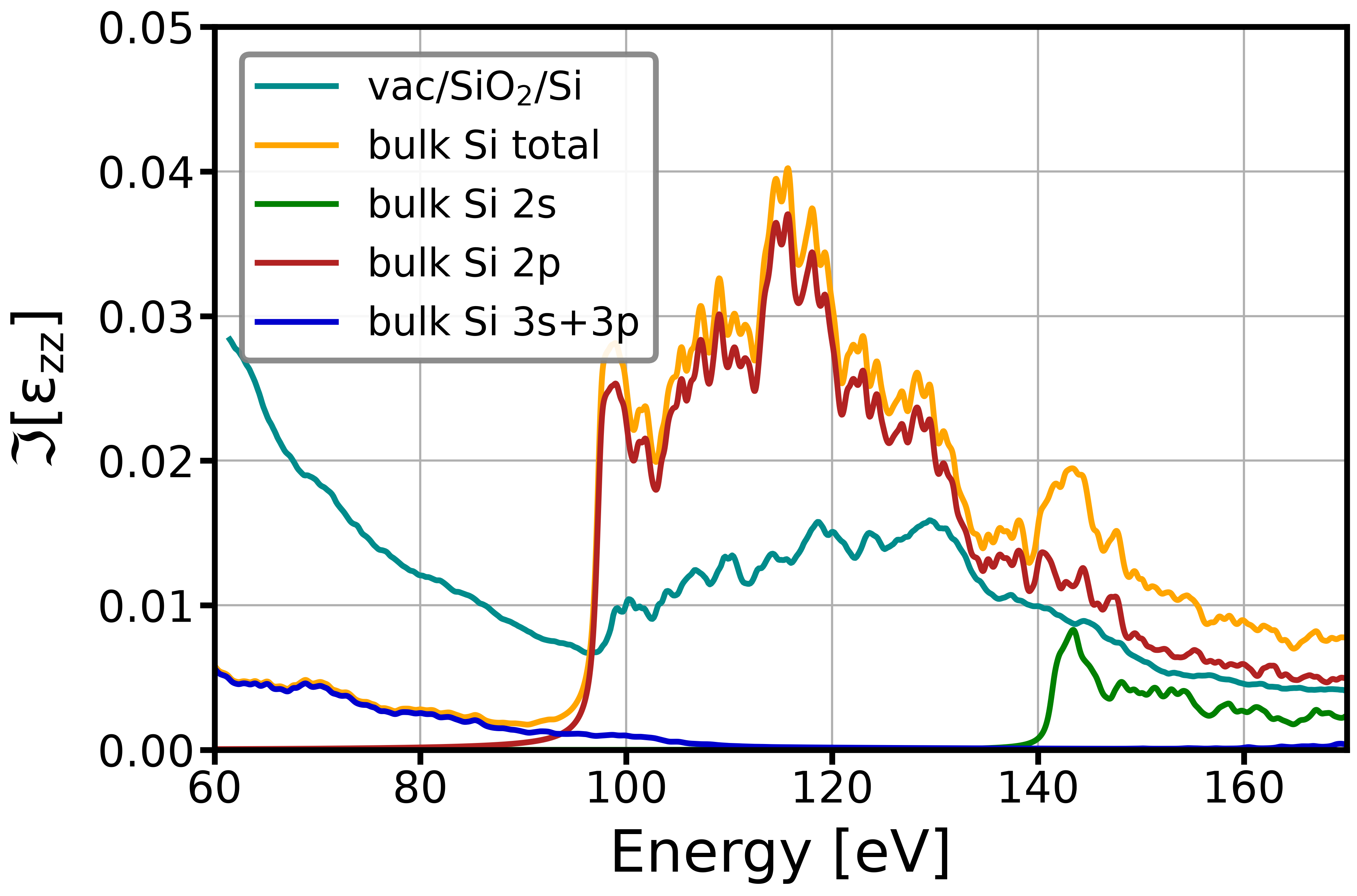}
  \end{minipage}%
  \begin{minipage}[c]{0.25\textwidth}
    \centering
    \makebox[\linewidth][l]{\textbf{B}}\\[0.5mm]
    \includegraphics[width=\linewidth]{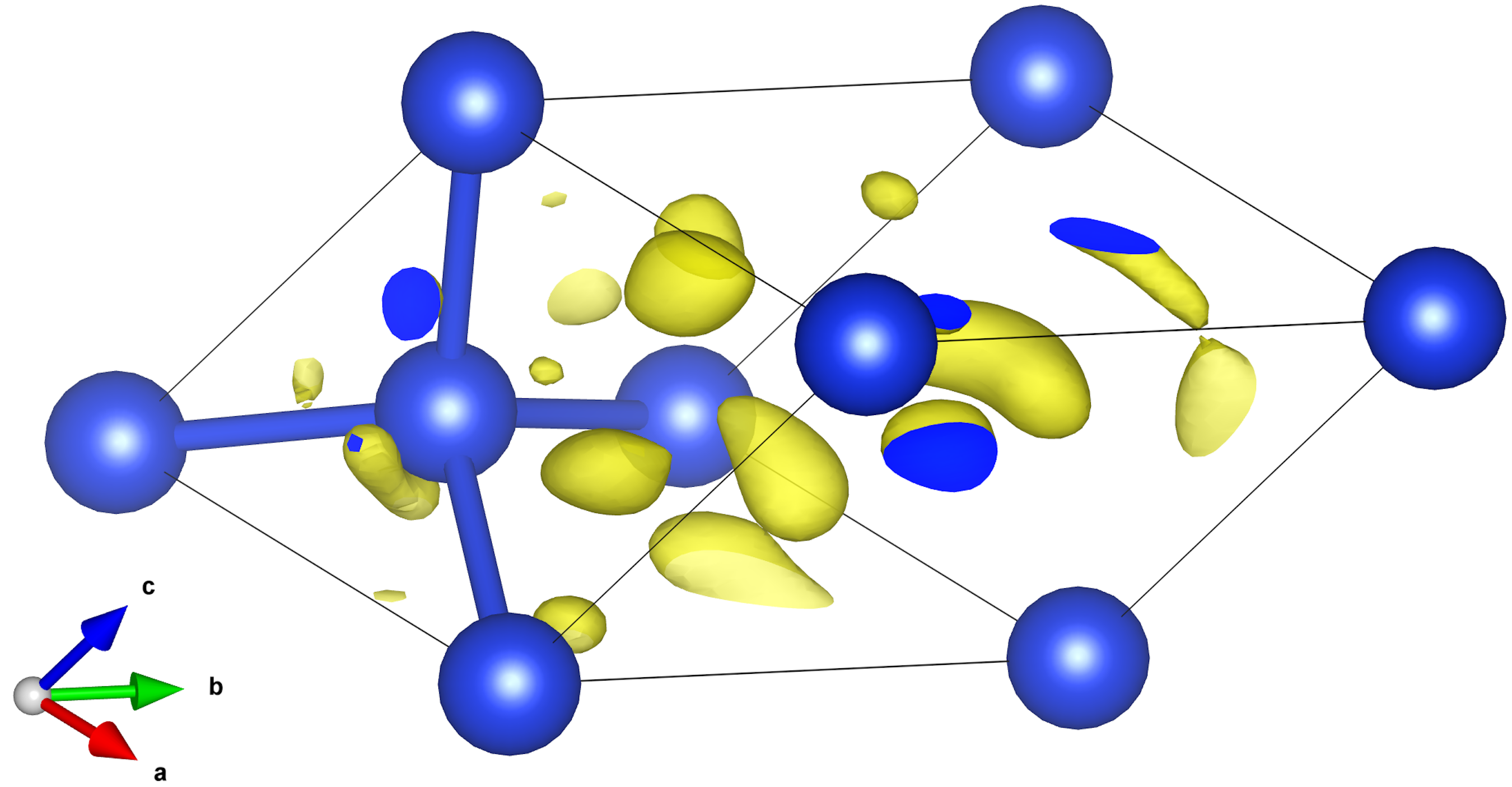}\\[1mm]
    \makebox[\linewidth][l]{\textbf{C}}\\[0.5mm]
    \includegraphics[width=\linewidth]{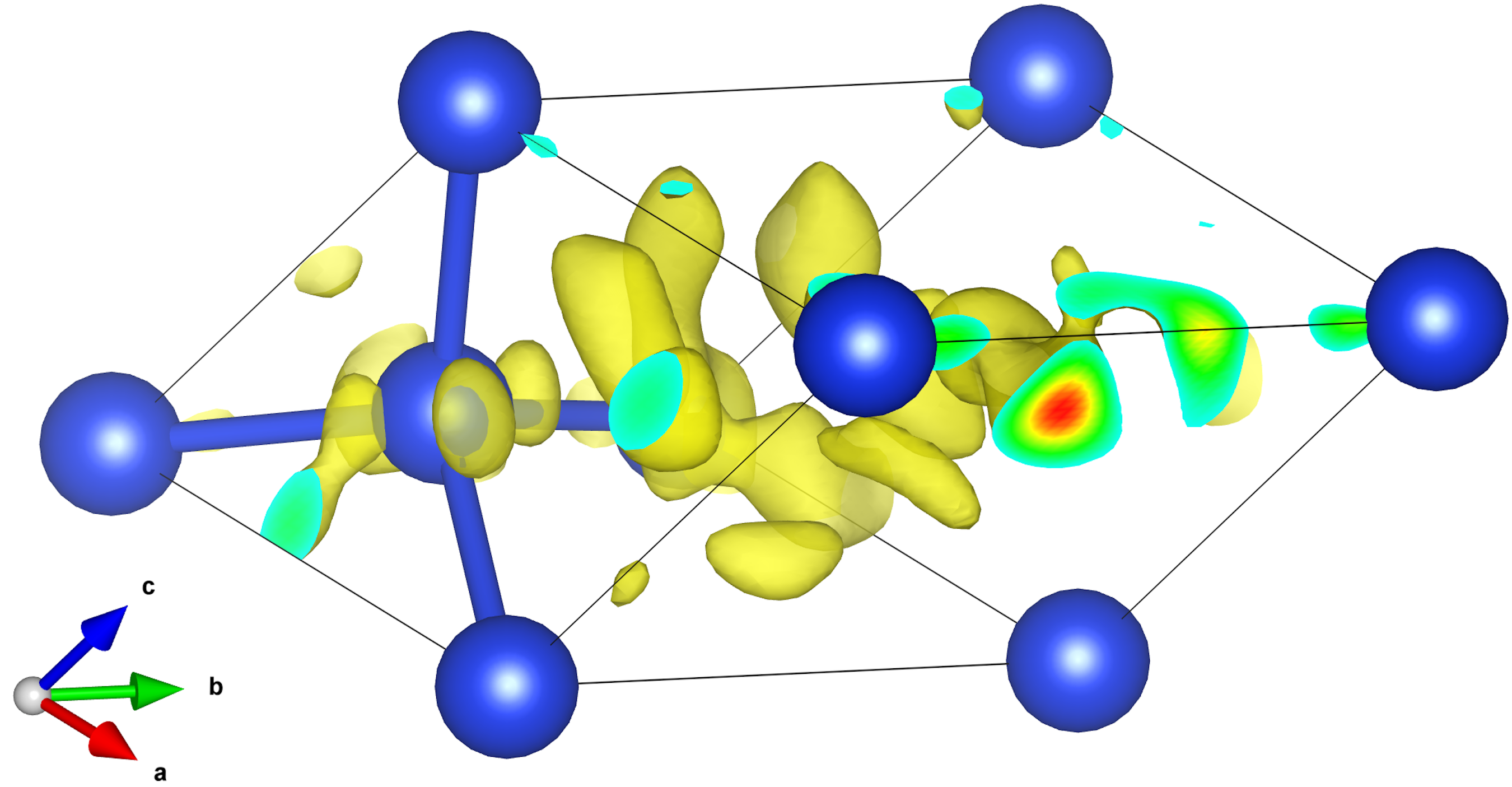}\\[1mm]
    \makebox[\linewidth][l]{\textbf{D}}\\[0.5mm]
    \includegraphics[width=\linewidth]{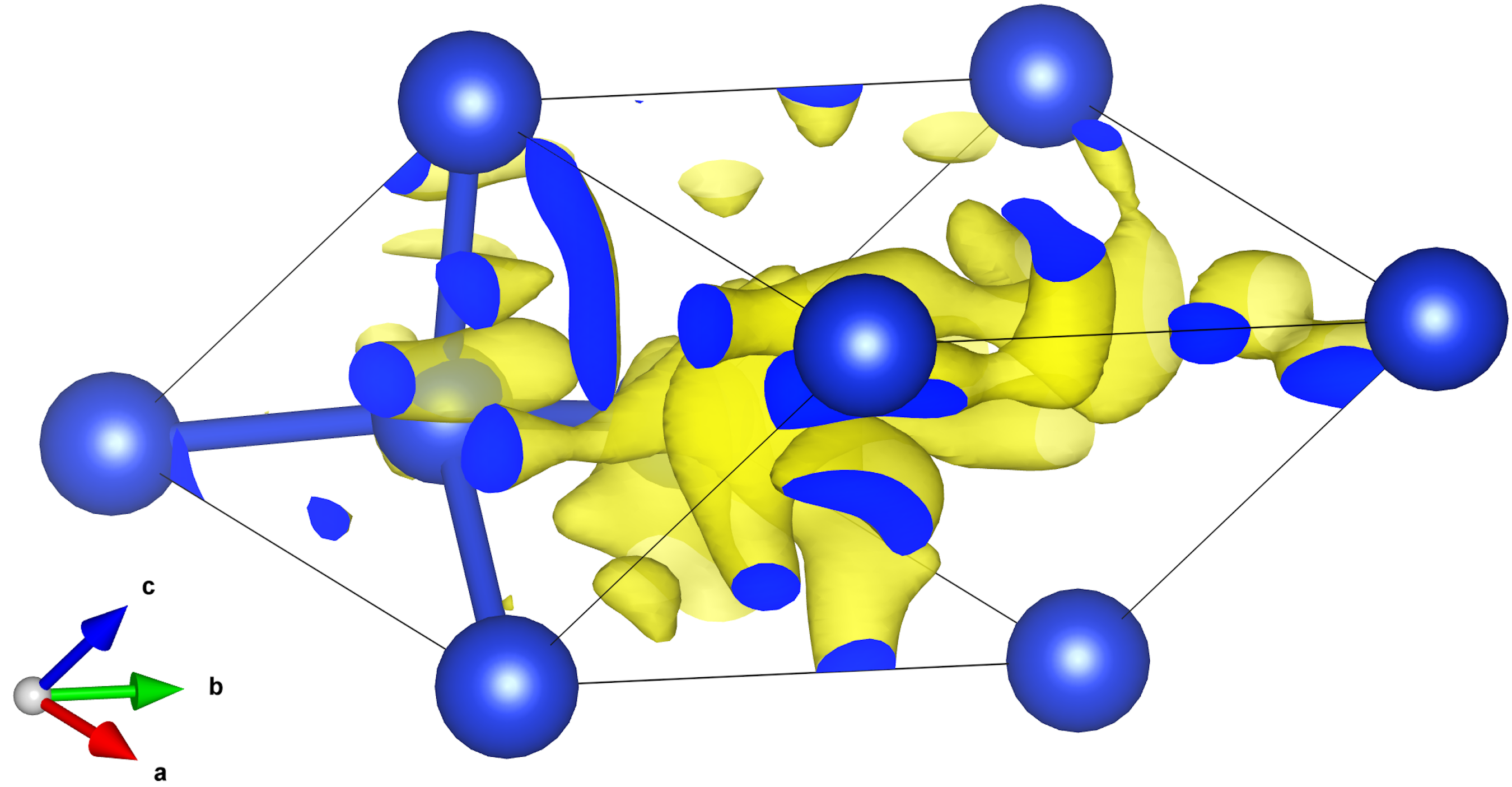}
  \end{minipage}
  \caption{\textbf{A} shows the imaginary part of the zz components of the dielectric function of a vacuum/SiO$_2$/Si slab (teal) and bulk Si (orange). Bulk Si's dielectric function is further broken down by possible initial states: $2s$ (green), $2p$ (red) and $3s$ and $3p$ (blue). \textbf{B}, \textbf{C}, and \textbf{D} are the densities of the electron states of the three most intense excitons from bulk Si $2p$ states between 140--160\,eV.}
  \label{fig:Figurespecandstates}
\end{figure*}

As the nature of the states involved in the XUV SHG process is not directly discernible from the simulation methodology used here, a time-reversal symmetry argument can be made for looking at the electron states from linear absorption as the candidate states for the XUV SHG.\cite{hoffman2025surface}
A random phase approximation calculation of the dielectric function of a vacuum/SiO$_2$/Si slab (teal, Figure~\ref{fig:Figurespecandstates}\textbf{A}) presents only a valence background in the 60--80\,eV direct resonance window, while Si $2p$ absorption extends through the half-resonance 140--160\,eV window.
As a direct Bethe-Salpeter equation (BSE) calculation of a slab (and the resulting excitons) is cost prohibitive with our approach, we focus instead on bulk Si.
The linear response in the direct resonance window (60--80\,eV) comes solely from the valence states in bulk Si (blue, Figure~\ref{fig:Figurespecandstates}\textbf{A}); the relevant excitations connecting to the Si $2p$ hole states responsible for the XUV SHG are therefore those at half-resonance (140--160\,eV).
The densities of the electron state of the three highest oscillator strength excitations in that range are shown on the right of Figure~\ref{fig:Figurespecandstates}.
While specific spectral causes are not simple to assign, one can tell that these are quite complicated transitions involving a large number of high-lying unoccupied states mixing together (plots in supplemental).
A very coarse breakdown is that the highest oscillator strength excitation (Figure~\ref{fig:Figurespecandstates}\textbf{B}) is associated with bands with $f$ character around the $\Gamma$ point, and the second and third highest oscillator strength excitations (Figures~\ref{fig:Figurespecandstates}\textbf{C} and \ref{fig:Figurespecandstates}\textbf{D}, respectively) are centered largely near the L point with bands that appear to have predominately $p$ type character, although the states are quite dense in that region.
A key takeaway from this analysis is that the densities of the electron state of these excitons are diffuse and mostly within the interstitial region between Si atoms, despite their respective hole states being quite atom localized. 
This large symmetry breaking between the spatial distribution of the electron and hole states is likely needed for a large cross section in these off-edge resonance phenomena.

In summary, XUV SHG studies at FELs have previously been performed under the assumption that linear phenomena provide a good predictor of energies where second order phenomena should also arise.
However, surprising and strong spectral features that do not align with a clear linear response are observed in the XUV SHG spectrum from a silicon photovoltaic heterojunction, showing that nonlinear spectra in the XUV region may be richer than previously known.
These features arise from the silicon dioxide/silicon interface due to the symmetry breaking and distortions at the interface bridging the bulk and oxide components, providing information about a buried interface that is not accessible by direct resonant XUV SHG at the Si $L_{2,3}$-edge.
We note that for other materials there may be other transitions which can be probed by nonlinear techniques which are largely inaccessible to linear probes and difficult to reach by direct resonance.

\section{Methods}

\subsection{Layered Junction Synthesis}

Layered solar cell junction slabs are fabricated using lightly doped $n$-type crystalline silicon wafer substrates ($>$100 $\Omega\cdot\mathrm{cm}$) with a very thin ($\sim$1\,nm) native SiO$_2$ layer on the surface. MoO$_3$ layers of approximately 4\,nm are deposited on the substrates by thermal evaporation. An indium tin oxide (ITO) layer of approximately 5\,nm is subsequently deposited by radio-frequency (RF) magnetron sputtering as a top coating to enable the device to function fully. The fabrication techniques have been detailed previously.\cite{Bullock2014,BULLOCK2015}
Thicknesses are verified by visible spectroscopic ellipsometry.

\subsection{XUV SHG Experimental Spectrum}

XUV SHG measurements are conducted at the soft x-ray FEL beamline BL1 of SACLA,\cite{Owada2018} which covers the photon energy range of $\hbar\omega=26$--150 eV. The FEL pulses have a width of $<$30 fs and energy per pulse of 75--135 $\mu$J. The energy resolution of the FEL is $\Delta E/E=\Delta(\hbar\omega)/\hbar\omega\sim$ 2\%. The FEL beam is focused to a spot size of 5.3 $\mu$m (horizontal) $\times$ 1.0 $\mu$m (vertical) on the solar cell slab sample using a Kirkpatrick-Baez mirror system. The FEL beam is incident on the sample surface with an angle 45$^\circ$. The incident FEL pulses are p-polarized, and the SH is detected without polarization analysis. The measured response therefore contains several $\chi^{(2)}$ tensor elements; we compare it with $\chi^{(2)}_{zzz}$.  Filters are used before the sample to remove some of the second harmonic from the source and to vary the incident pulse power to enable a determination of $\chi^{(2)}$. The experimental details of the setup have been discussed elsewhere, specifically the setup for solid samples \cite{horio2023detecting}, and the grating/detector system.\cite{Berger2021,Woodahl2023}

Due to the intensity needed to induce nonlinear effects, the photovoltaic stack sample is damaged by each FEL pulse. Therefore, the sample is rastered after each FEL shot so that each shot is incident on a fresh region of the sample. Each point in the experimental $|\chi^{(2)}|^2$ spectrum represents a few thousand FEL shots. Each shot is recorded on a 2D detector such that the FEL fundamental and SH are visible. The image for each shot is reduced to the region of interest around the fundamental and SH and summed vertically to generate a single-shot spectrum. For more precise monitoring of the FEL intensity with each shot we also record the voltage from a gas monitor. At a given photon energy, the single shot spectra are binned by the associated FEL fundamental intensity recorded on the gas monitor such that each bin contains 200 single-shot spectra. Each bin of single-shot spectra is then averaged and the SH peak is fit to a Gaussian. The Gaussian amplitude of the binned SH vs the average value of the FEL fundamental from the gas monitor for that bin is fit to a second order polynomial, whose linear term accounts for residual second harmonic of the FEL source reflected from the sample.\cite{Sumi2022,Sumi2023} The coefficient of the quadratic term is taken to be the $|\chi^{(2)}|^2$ response for a specific photon energy. This process is repeated with different sets of filters to vary the FEL intensity and the average and standard deviation of $|\chi^{(2)}|^2$ between these runs is represented in Fig.~\ref{fig:SHGspectrum} for each photon energy measured.

\subsection{Theoretical Model}

While many structural models for the Si/SiO$_2$ interface have been proposed,\cite{PhysRevLett.59.213,PhysRevB.41.5061,PhysRevLett.74.1024,PhysRevB.53.10942,Lenosky1999Energetics,PhysRevLett.85.1298} Si/SiO$_2$ slab structures are constructed here following \citet{PhysRevB.63.115314} where $\alpha$-quartz is the preferred oxide structure when it is $>$1\,nm thick. 
The slab consists of a bulk of 4 Si cubic unit cells, a Si monolayer bridging the bulk, and 8 (Si) monolayers of $\alpha$-quartz strained to match Si's lattice constant (5.4307\,\AA).
The oxide is passivated with silanols, and the back face of the bulk is passivated with hydrogen to suppress unphysical surface effects.
When indicated, a single bilayer of $\alpha$-MoO$_3$~\cite{kihlborg1963least} strained to match the Si lattice constant is placed $\sim$1.2\,\AA{} above the oxide. 
At least 15\,\AA{} of vacuum separates the two sides of the slab.
All atoms but those of the bulk Si are allowed to relax, keeping the lattice fixed.
The optimizations are performed with VASP 5.4.4 using the SCAN\cite{Sun2015} meta-GGA functional and the -D3 dispersion correction with Becke-Johnson damping\cite{D3BJ,Brandenburg2016}.
The optimizations use a $4\times4\times1$ $\Gamma$-centered $k$-grid, 400 eV plane-wave cutoff, $10^{-6}$ eV energy threshold, and $10^{-2}$ eV/\AA{} force threshold.
The projector augmented wave (PAW)\cite{Blochl1994} pseudopotentials have valence configurations: Si - $3s^23p^2$, O - $2s^22p^4$, Mo - $4p^65s^24d^4$, and H - $1s^1$.

Following the approach of previous studies of soft X-ray SHG,\cite{lam2018soft} the second-order susceptibility tensor components normal to the slab termination, $\chi_{zzz}^{(2)}$, are calculated using the second-order response formalism of \citet{Sharma2004} as implemented in the all-electron exciting~\cite{Gulans2014} neon-21 package.
The PBE~\cite{Perdew1996} GGA functional is used for these calculations along with a dipole correction along the $c$-axis and a $7\times7\times1$ $\Gamma$-centered $k$-grid.
Only the Si 2s and 2p electrons are moved from the core to the valence versus exciting's default species definition, and the muffin tin radii (a.u.) are Si - 2.00, O - 1.02, Mo - 2.00, and H - 0.75.
5000 empty Kohn-Sham states are added to the simulations.
The exciting code is modified to restrict the initial states in the second-order SHG response to enable extracting the susceptibility from only Si $2s$, Si $2p$, Mo $4s$, and Mo $4p$ states as well as to only obtain the response from specific atomic layers.
As the slabs we construct here naturally break centrosymmetry, no Si $2s$ or $2p$ states are kept in core like was done in \citet{lam2018soft}.
Keeping states in core excludes them from the calculation to construct an effective inversion symmetry breaking between the top and bottom surfaces of a slab.
Partitions are selected based on the delocalization of the Si $2p$ Kohn-Sham states at the $\Gamma$-point.
As with the previous study,\cite{lam2018soft} a heuristic energy dependent imaginary value, $\varepsilon_m + \sfrac{i}{\lambda}\sqrt{\varepsilon_m - E_F}$,\cite{Klinkusch2009} is added to the unoccupied Kohn-Sham states in the SHG calculation to account for lifetime broadening and mitigate numerical noise in the response function.
In this expression, E$_F$ is the Fermi energy and $\lambda$ is the inelastic mean-free path of electrons~\cite{Sorini2006} which is taken as 5.3\,\AA{} for bulk Si at the experimental photon energies.\cite{Powell2010NIST}
The dielectric functions of the slabs are determined using the Dyson equation with the random phase approximation exchange-correlation kernel as implemented in exciting;\cite{PhysRevLett.88.037601} due to computational constraints, those slabs contain only half as many bulk and oxide Si monolayers.
Linear absorption spectra and excitons of bulk Si are calculated using the Bethe-Salpeter equation as implemented in exciting.\cite{10.1039/b903676h}
Since PBE underestimates the $L_{2,3}$ edge, a rigid shift of 7.065\,eV is applied to the simulated curves to align them with experiment.

\section{Data availability}

The data that support the findings of this study are included in the manuscript. Additional raw data will be made available via Dryad upon publication.

\section{Author Contributions}

\textbf{B.R.N.}: Conceptualization, Methodology, Software, Validation, Formal Analysis, Investigation, Writing - Original Draft, Writing - Review and Editing, Visualization;
\textbf{S.W.D.}: Conceptualization, Methodology, Software, Validation, Investigation, Writing - Review and Editing;
\textbf{F.B.}: Conceptualization, Methodology;
\textbf{A.R.}: Investigation, Methodology, Resources, Writing - Review and Editing;
\textbf{J.B.}: Conceptualization, Methodology, Resources, Supervision, Project Administration, Funding Acquisition;
\textbf{D.R.S.}: Methodology, Software, Investigation;
\textbf{J.D.K.}: Conceptualization, Supervision;
\textbf{Y.K.}: Methodology, Software, Investigation, Resources;
\textbf{S.O.}: Methodology, Software, Investigation, Resources;
\textbf{T.S.}: Methodology, Software, Investigation, Resources;
\textbf{X.S.}: Methodology, Software, Investigation, Resources;
\textbf{I.M.}: Conceptualization, Resources, Data Curation, Supervision, Project Administration, Funding Acquisition;
\textbf{K.V.L.}: Conceptualization, Methodology, Software, Validation, Formal Analysis, Investigation, Resources, Data Curation, Writing - Original Draft, Writing - Review and Editing, Visualization, Supervision;
\textbf{W.S.D.}: Conceptualization, Methodology, Investigation, Data Curation, Writing - Review and Editing, Supervision, Project Administration, Funding Acquisition;
\textbf{M.W.Z.}: Conceptualization, Methodology, Investigation, Data Curation, Writing - Original Draft, Writing - Review and Editing, Supervision, Project Administration, Funding Acquisition;
\textbf{C.P.S.}: Conceptualization, Methodology, Investigation, Data Curation, Writing - Original Draft, Writing - Review and Editing, Supervision, Project Administration, Funding Acquisition

\section{Competing interests}

The authors declare no competing interests.

\begin{acknowledgments}
This work was supported by the U.S. Department of Energy, Office of Science, Basic Energy Sciences, Chemical Sciences, Geosciences, and Biosciences Division under Contract
DE-SC0023355. This research used resources of the National Energy
Research Scientific Computing Center, a DOE Office of Science User Facility supported
by the Office of Science of the US DOE, under contract no. DE-AC02-05CH11231, using NERSC award BES-ERCAP0031431 and BES-ERCAP0035240. This work also used
the Expanse supercomputer at the San Diego Supercomputing Center through allocation
PHY210131 from the Advanced Cyberinfrastructure Coordination Ecosystem: Services \&
Support (ACCESS) program, which is supported by NSF Grant No. 2138259, No. 2138286,
No. 2138307, No. 2137603, and No. 2138296. B.R.N. acknowledges funding from the National Science Foundation Graduate Research Fellowship Program (DGE 1752814).  B.R.N. further acknowledges funding by the UC Office of the President within the Multicampus Research Programs and Initiatives (M23PR5931). W.S.D. acknowledges support from the U.S. Department of Energy, Office of Science, Basic Energy Sciences, Chemical Sciences, Geosciences, and Biosciences Division under FWP  FP00014920. M.Z. acknowledges funding by the U.S. Department of Energy (DE-SC0024123), the Rose Hills Foundation, the Camille and Henry Dreyfus Foundation, and the Hellman Foundation.  These experiments were performed at SACLA BL1 with the approval of the Japan Synchrotron Radiation Research Institute (JASRI) (Proposal Nos. 2023B8048, 2024A8050, and 2026A8046)
\end{acknowledgments}

\bibliography{REFERENCES}

\end{document}